\documentclass{article}
\usepackage[version=3]{mhchem} 
\usepackage{siunitx} 
\DeclareSIUnit{\litre}{l}
\usepackage{physics}
\usepackage{physunits}
\usepackage[margin=0.5in]{geometry}
\usepackage{graphicx}

\begin{document}

\title{Large-Scale Bottom-Up Fabricated 3D Nonlinear Photonic Crystals}

\maketitle

\author{Dr. Viola Valentina Vogler-Neuling}\footnote[1]{ETH Zurich, Department of Physics, Institute for Quantum Electronics, Optical Nanomaterial Group, HPT H 2, Auguste-Piccard-Hof 1, CH-8093 Zurich, Switzerland, E-mail: grange@phys.ethz.ch},\footnote[2]{University of Fribourg, Adolphe Merkle Institute, Soft Matter Physics Group, Chemin des Verdiers 4, CH-1700 Fribourg, Switzerland, E-mail:viola.vogler-neuling@unifr.ch}
\author{Ülle-Linda Talts}\footnotemark[1]
\author{Rebecca Ferraro}\footnotemark[1]
\author{Helena Weigand}\footnotemark[1]
\author{Giovanni Finco}\footnotemark[1]
\author{Joel Winiger}\footnote[3]{ETH Zurich, Department of Information Technology and Electrical Engineering, Institute for Electromagnetic Fields, ETZ K 81, Gloriastrasse 35, CH-8092 Zurich, Switzerland}
\author{Dr. Peter Benedek}\footnote[4]{ETH Zurich, Department of Information Technology and Electrical Engineering, Institute for Electronics, ETZ H 96, Gloriastrasse 35, CH-8092 Zurich, Switzerland}
\author{Dr. Justine Kusch}\footnote[5]{ETH Zurich, ScopeM, HPM C 52.1, Otto-Stern-Weg 3, CH-8093 Zurich, Switzerland}
\author{Dr. Artemios Karvounis}\footnotemark[1]
\author{Prof. Vanessa Wood}\footnotemark[4]
\author{Prof. Jürg Leuthold}\footnotemark[3]
\author{Prof. Rachel Grange}\footnotemark[1]

\begin{abstract}
Nonlinear optical effects are used to generate coherent light at wavelengths difficult to reach with lasers. Materials periodically poled or nanostructured in the nonlinear susceptibility in three spatial directions are called 3D nonlinear photonic crystals (NPhCs). They enable enhanced nonlinear optical conversion efficiencies, emission control, and simultaneous generation of nonlinear wavelengths. The chemical inertness of efficient second-order nonlinear materials ($\chi^{(2)}$)  prohibited their nanofabrication until 2018. The current method is restricted to top-down laser-based techniques limiting the periodicity along $z$-axis to $\SI{10}{\micro\meter}$. We demonstrate the first bottom-up fabricated 3D NPhC in sol-gel derived barium titanate by soft-nanoimprint lithography: a woodpile with eight layers and periodicities of $\SI{1}{\micro\meter}$ ($xy$-plane) and $\SI{300}{\nano\meter}$ ($z$-plane). The surface areas exceed $5.3\cdot 10^4~\SI{}{\micro\meter\squared}$, which is two orders of magnitude larger than the state-of-the-art. This study is expected to initiate bottom-up fabrication of 3D NPhCs with a supremely strong and versatile nonlinear response.
\end{abstract}

\section*{Introduction}
Linear photonic crystals have been demonstrated on different material platforms in one, two, and three dimensions.\cite{Butt2021,Shen2016,Yue2015,Wen2008,Cersonsky2021} Their applications range from photonic integrated circuits \cite{Anand04}, solar cells \cite{Liu2019}, topological photonic crystals \cite{Ozawa2019,Lu2014} to optical sensing \cite{Zhang2015}. However, the emitted spectrum of linear photonic crystals is limited to incident wavelengths. Their application range can be significantly expanded if the photonic crystals are made out of a nonlinear optical material, for example, a nonlinear optical node, i.e. a photonic crystal structure that emits light at different wavelengths along different crystalline axes.\cite{Keren-Zur2018}\\
Nonlinear optical materials enable light-matter interaction in various ways, e.g., by generating different wavelengths, entangled photons, etc.\cite{Boydbook2003} Among the nonlinear optical materials, the second-order optical materials ($\chi^{(2)}$) giving rise to second-harmonic generation (SHG) have the highest nonlinear responses.\cite{VoglerNeuling22} A non-zero susceptibility in the second-order term $\chi^{(2)}$ is only present in non-centrosymmetric materials, which significantly reduces the number of available platforms.\cite{Boydbook2003} The challenge in the engineering of solution-derived $\chi^{(2)}$-materials compared to their linear counterparts is that the material has to be crystalline or at least polycrystalline without exhibiting inversion symmetry.\cite{VoglerNeuling22, Savo2020} Even though $\chi^{(2)}$-materials have the highest nonlinear conversion efficiencies ($P_{\mathrm{SHG}}/P_{\mathrm{Fundamental}}$), these efficiencies are still intrinsically weak.\cite{Fejer1994}\\
By combining a 3D linear photonic crystal geometry with a nonlinear optical material, enhanced nonlinearities can be achieved at the band edges of the photonic bandstructure where the group velocity ($v_{\mathrm{g}}=d\omega/dk$) is approaching zero. A group velocity approaching zero allows for enhanced light-matter interaction, an increased phase shift enabling a smaller device footprint, reduced operational power, and more compressed laser pulses.\cite{Soljacic2002,Figotin2006,Pedersen2008}\\ 
Dielectric nanostructures, periodic in the nonlinear susceptibility in all three spatial directions, are called 3D nonlinear photonic crystals (NPhCs). 3D NPhCs can be used as a nonlinear optical node or efficient source for different wavelengths. The achievable nonlinear intensities in the quasi-phase matching regime scale with the length of photonic crystals $I_{\mathrm{SHG}}\propto L_{\mathrm{c}}$.\cite{Zhang2021} Therefore, larger crystals will lead to stronger nonlinear signals and high-quality photonic crystal cavities could be engineered.\cite{Lin2001}\\
Since the theoretical prediction of 3D NPhCs in 1998 \cite{Berger1998}, there was no experimental technique to modulate the $\chi^{(2)}$ in three spatial dimensions due to the chemical inertness of $\chi^{(2)}$-materials until 2018 when two research groups experimentally realized the first 3D NPhCs in \ce{LiNbO3} and $\mathrm{Ba}_{0.77}\mathrm{Ca}_{0.23}\mathrm{Ti}\mathrm{O}_3$ (BCT).\cite{VoglerNeuling22,Wei2018,Xu2018}\\
These breakthrough works used a material processing technique based on ultrafast light pulses by periodically lowering the local nonlinearity due to partial destruction of the crystalline structure at the beam's focal point or by periodic inversion of ferroelectric domains due to spontaneous polarization under sub-damage threshold exposure.\cite{Zhang2021} Laser-based processing methods are restricted in the critical dimensions of the structures. In the $xy$-plane (parallel to the substrate), the structures are limited by the focus area of the laser and in the $z$-direction (perpendicular to the substrate) by the Rayleigh length of the laser beam which restricts to periodicities of above $\SI{10}{\micro\meter}$.\cite{Wei2018} The demonstrated surface area is also limited. So far, recent experiments did not exceed three layers, around 500 unit cells, and the smallest periodicity is limited to $\SI{500}{\nano\meter}$ in $xy$-plane.\cite{Wei2018,Liu2020,Xu2022} Following the same approach, four different 3D nonlinear photonic crystal with cubic unit cell structures in $z$-cut $\mathrm{Sr}_{0.61}\mathrm{Ba}_{0.39}\mathrm{Nb}_2\mathrm{O}_6$ were realized.\cite{Chen2022} Another fabrication method is growing the 3D nonlinear photonic crystals with a Czochralski method. 3D nonlinear photonic crystals by this method have been demonstrated in potassium-tantalate-niobate (KTN) \cite{Li2020} and BCT crystals \cite{Xu2016}. This technique can generate large surface areas on the order of several $\SI{}{\milli\meter\squared}$ but with less control over the crystalline orientation and nano structuration. In the field of 3D nonlinear photonic crystals, the major technical challenge is to fabricate high-precision large-scale 3D nonlinear photonic crystals for efficient nonlinear conversion efficiencies.\cite{Zhang2021} \\ 
In this work, to overcome the limitations of typical top-down optical processing and increase topology diversity, we fabricate woodpile photonic crystal structures by soft-nanoimprint lithography (SNIL). The critical dimensions of this technique are determined by the minimal structures achievable with electron-beam lithography in silicon, which acts as a master mold.\cite{Einck2021} The woodpile photonic crystal structure, known to be one of the 3D photonic structures exhibiting a full linear photonic band gap, is favorable over 3D network structures with a complete bandgap such as the rod-connected diamond structure due to its compatibility with layer-by-layer fabrication.\cite{Joannopoulos2007,Maldovan2004} A large variety of linear woodpile photonic crystals have been demonstrated with direct laser writing and soft-nanoimprint lithography.\cite{Beaulieu2014,Fischer2011} While direct laser writing is a powerful technique for structuring photopolymerizable linear materials and recently the achievable periodicity limit was decreased to below $\SI{150}{\nano\meter}$ \cite{Hahn2021}, combining the use of femtosecond (fs) lasers with a nonlinear composite matrix poses a challenge of parasitic SHG generation and results in uncontrolled crosslinking of the surrounding material.\\
Using soft-nanoimprint lithography together with a barium titanate $\ce{BaTiO3}$ sol-gel chemistry, we thus demonstrate the first bottom-up fabricated 3D nonlinear photonic crystal in a woodpile geometry. Following this approach, we are increasing the number of layers in $z$-direction to 8. We are decreasing the periodicity in $z$-direction to only $\SI{300}{\nano\meter}$ with $xy$-periodicities of $\SI{1}{\micro\meter}$. The surface areas of our crystals yield more than $5.3\cdot 10^{4}~\SI{}{\micro\meter\squared}$ corresponding to $5.3\cdot 10^4$ unit cells in $xy$-plane. This is two orders of magnitude higher than the state-of-the-art.\\
We anticipate our study to be a starting point for bottom-up fabrication of 3D NPhCs and for developing sol-gel derived materials with optimized nonlinear responses. 3D nonlinear photonic crystals are the ideal testbed to study new fundamental physics and to demonstrate multi-wavelength coherent nonlinear light sources as well as compact sources for entangled photons.\cite{Keren-Zur2018}

\section*{Results}
\subsection*{Nanofabrication of Large-Scale 3D Nonlinear Photonic Crystals by Soft-Nanoimprint Lithography}
We introduce an alternative fabrication route to established laser-based techniques for 3D NPhCs, namely soft-nanoimprint lithography of barium titanate prepared by sol-gel chemistry. In our previous work, we imprinted barium titanate nanoparticles of a diameter of $\SI{50}{\nano\meter}$ together with a titanium dioxide sol-gel into a woodpile photonic crystal geometry.\cite{Vogler-Neuling2020} While nanoparticles with diameters above $\SI{42}{\nano\meter}$ are known to exhibit bulk second-harmonic generation, they induce structural roughness and cracks in the woodpile, and thus induce unwanted incoherent scattering.\cite{Kim2013a}
\begin{figure}[h!]
  \centering
  \includegraphics[scale=0.75]{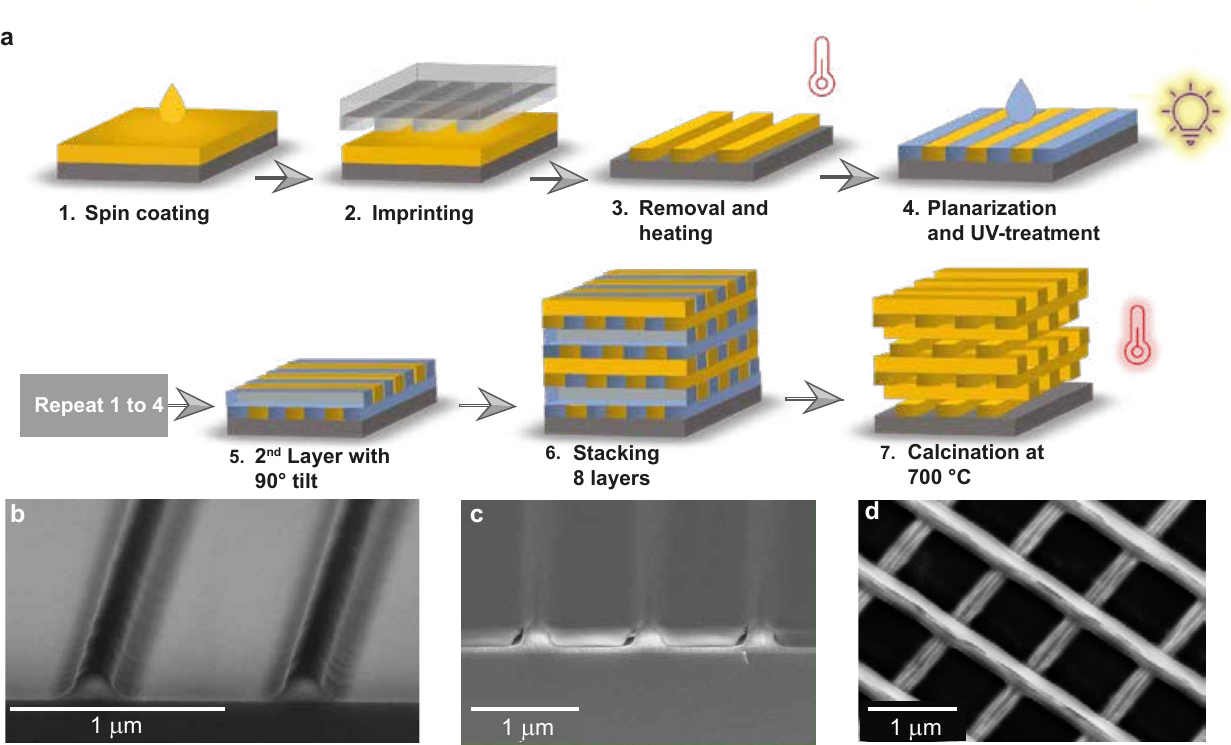}
  \caption{\textbf{a} Fabrication flow diagram for woodpile photonic crystals. After step 2, the sample is heated at $\SI{60}{\celsius}$ for $\SI{3}{\hour}$. \textbf{b} SEM cross-section image of one layer showing no residual layer between the silicon substrate and the structures. The dimensions of a single imprinted lathe yer are $\SI{1}{\micro\meter}$ for unit cell periodicity, rod height of $\SI{150}{\nano\meter}$ and width of $\SI{185}{\nano\meter}$. \textbf{c} SEM image showing the sample after planarization of the first layer with NOA60 negative photoresist. \textbf{d} SEM image of a two-layer sample showing that the rods are free-standing after calcination at $\SI{700}{\celsius}$.}
  \label{Fig.1}
\end{figure}
 In this work, we follow the same fabrication flow (Figure \ref{Fig.1}a).\cite{Vogler-Neuling2020} To increase the homogeneity of our structures. However, we exchange the barium titanate nanoparticles with a barium titanate sol-gel synthesized from barium acetate, acetic acid, titanium(IV)isopropoxide, and acetylacetone following the procedure described in \cite{Edmondson2020} (Methods).
A silicon master mold of a grating with a period of $\SI{1}{\micro\meter}$, a height of $\SI{340}{\nano\meter}$ and a width of $\SI{500}{\nano\meter}$ is fabricated by standard cleanroom techniques and polydimethylsiloxane (PDMS) daughter molds are prepared (Methods). 
 $\SI{10}{\micro\litre}$ of barium titanate sol-gel dissolved in $\SI{10}{\micro\liter}$ anhydrous methanol is spin coated on either fused quartz (for nonlinear optical characterization) or silicon substrates (for nanostructure characterization) at a velocity of $\SI{1000}{\rpm}$ and an acceleration of $\SI{1000}{\rpm\per\second}$ for $\SI{5}{\second}$. The main challenge with this procedure lies in engineering the processing parameters such that there is no residual BTO layer after the imprinting (Figure \ref{Fig.1}b). The parameters with the most significant influence are the ratio between barium titanate sol-gel and anhydrous methanol, the total amount of liquid, and the velocity of the spin coating. Directly after spin coating, the PDMS mold is placed onto the substrate, pressed down with a $\SI{35}{\gram}$ weight, and the solvents evaporate through the PDMS mold at $\SI{60}{\celsius}$ for three hours. Afterward, the PDMS is removed, and the sample is slowly heated to $\SI{400}{\celsius}$ to prevent dissolving during the stacking of the next layer.
 The imprinted grating structure has a period of $\SI{1}{\micro\meter}$, a height of $\SI{150}{\nano\meter}$ and a width of $\SI{185}{\nano\meter}$ (Figure \ref{Fig.1}b). This corresponds to a duty cycle of $\SI{18.5}{\percent}$. To stack the next layer, the grating was planarized with $\SI{10}{\micro\liter}$ of a $15~\mathrm{wt.}~\SI{}{\percent}$ solution of negative photoresist (NOA60) and exposed to UV light achieving an exact coverage of the imprinted structures (Methods and Figure \ref{Fig.1}c). After processing, the samples are high-temperature annealed to $\SI{700}{\celsius}$ for $\SI{10}{\minute}$ with a heating and cooling rate of $\SI{2}{\celsius\per\minute}$ to transform the sol-gel into polycrystalline and partially non-centrosymmetric barium titanate and to make the structures free-standing by decomposing the planarization photoresist (Figure \ref{Fig.1}d).
 \begin{figure}[h!]
 \centering
  \includegraphics[scale=0.75]{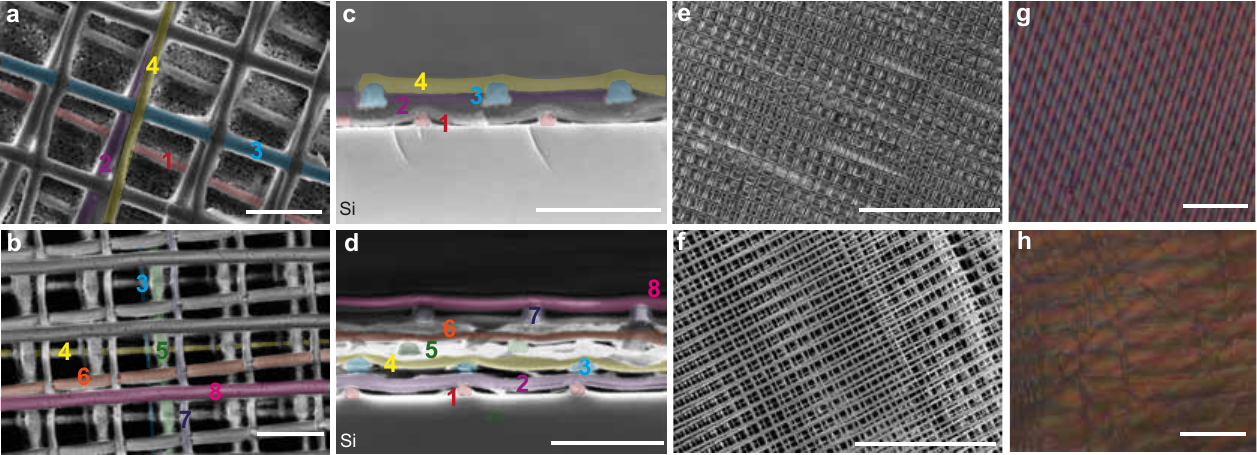}
  \caption{SEM images of \textbf{a} four and \textbf{b} eight layers of the fabricated barium titanate woodpile photonic crystals after annealing. The focal depth of the SEM was not high enough to resolve all the eight layers in \textbf{b} from the top. \textbf{c} and \textbf{d} show the corresponding cross-sectional SEM images of the structures, confirming the eight layers. The scale bars in \textbf{a}-\textbf{d} are $\SI{1}{\micro\meter}$. \textbf{e} and \textbf{f} SEM images of a larger surface area of the four-and eight-layer sample. Scale bars are $\SI{10}{\micro\meter}$. \textbf{g} and \textbf{h} are optical microscopy images of the same samples. Whereas the four-layer \textbf{g} sample is defect-free on this length scale, the \textbf{h} eight-layer sample shows a few areas of defects. Scale bars are $\SI{40}{\micro\meter}$. A Moiré pattern is visible in \textbf{g} and \textbf{h} from the small rotation imperfections.}
  \label{Fig.2}
\end{figure}
 Following this procedure, one to eight layers of a woodpile photonic crystal were fabricated by rotating the daughter mold for every layer by $\SI{90}{\degree}$ (Figure \ref{Fig.1}b,d). Figure \ref{Fig.2}a shows a four-layer woodpile structure, and Figure \ref{Fig.2}b an eight-layer woodpile sample. To image the layered nature of our samples, we cleaved the structures and imaged the cross-sections in the SEM (Figures \ref{Fig.2}c and d). The images reveal the four and eight layers of the photonic crystals, respectively. The simplest woodpile geometry consists of a log orientation A and an orthogonal orientation B, called ABAB stacking geometry. The more sophisticated version, ABCD, has two additional layers, C and D, with the same orientation as A and B but with an offset by half of the horizontal spacing (Figure S2).\cite{Joannopoulos2007} Our structures follow an ABAB stacking geometry rather than ABCD. Slight rotation errors due to PDMS mold positioning by hand in the range of $\pm \SI{5}{\percent}$ occurred (SI Figure S1). Simulation results show that the stacking, as well as the small rotation errors have a minor effect on the expected band gap position due to the connectivity of the dielectric network (SI Figure S2).\cite{Sigalas1999} SEM images in Figures \ref{Fig.2}e and f show large defect-free surface areas on the order of $\SI{400}{\micro\meter\squared}$ in the $xy$-plane. In the $z$-plane, four unit cells are present for the eight-layer sample. Optical microscopy images (Figures \ref{Fig.2}g and h) reveal even larger surface areas on the order of $2.1\cdot10^{4}~\SI{}{\micro\meter\squared}$ this corresponds to $2.1\cdot10^4$ unit cells in the $xy$-plane. At this scale, the four-layer sample is still completely defect-free, whereas a few defects become apparent for the eight-layer sample. Furthermore, due to the small shifts between consecutive layers, a Moiré pattern can be observed.\\
 In summary, we can state that we fabricated the first bottom-up woodpile photonic crystals out of a nonlinear optical material with up to eight-layers, periods of $\SI{1}{\micro\meter}$ in $xy$-plane, periods of $\SI{300}{\nano\meter}$ in $z$-plane and $2.1\cdot10^4$ unit cells.

\subsection*{Characterization of Nonlinear Optical Material}
\begin{figure}[h!]
 \centering
  \includegraphics[scale=0.75]{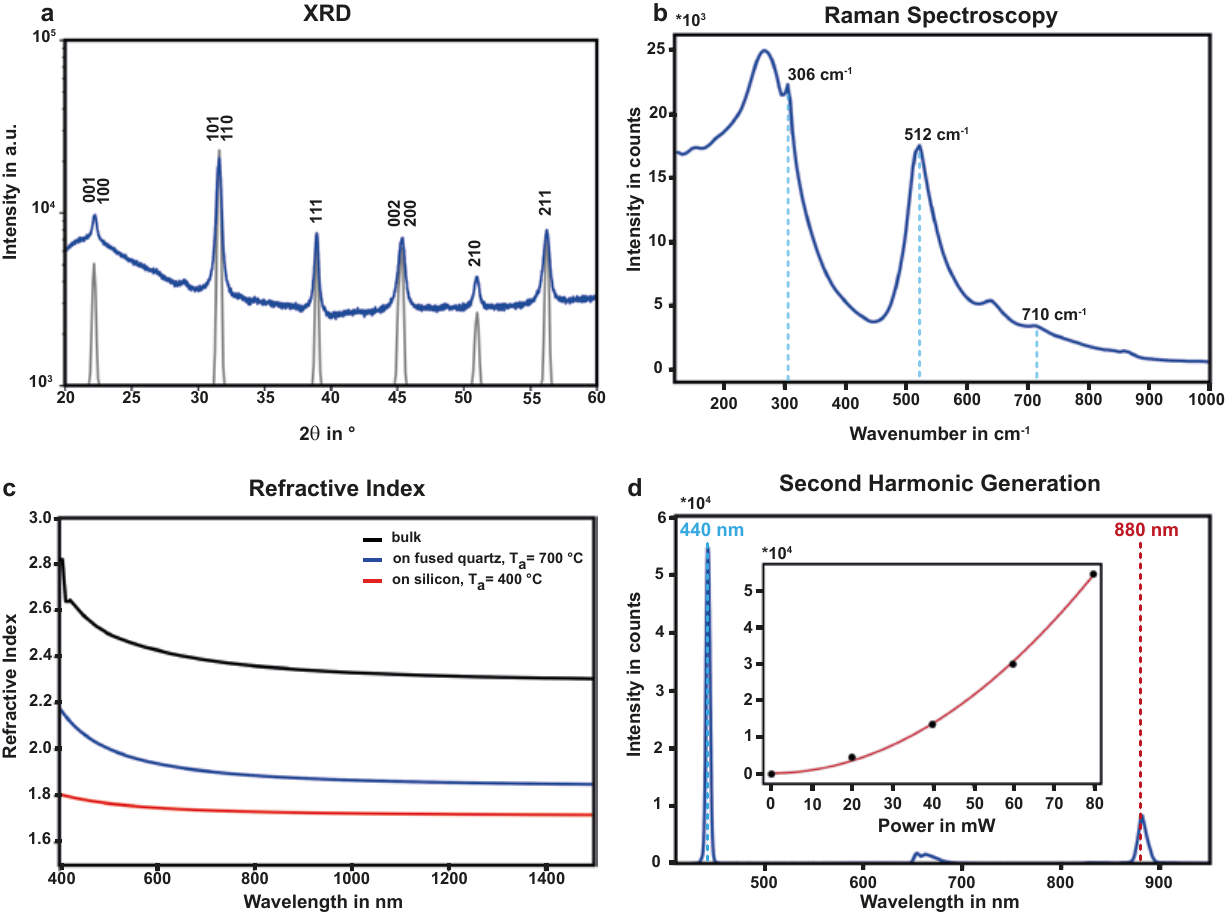}
  \caption{\textbf{Characterization of sol-gel derived barium titanate} \textbf{a} XRD-measurement on an annealed thin film of barium titanate on fused quartz. In blue is the measured data, and in grey is the corresponding ICSD reference for tetragonal barium titanate (ICSD-67520). \textbf{b} Raman spectroscopy measurement on the same barium titanate thin film demonstrating a tetragonal ferroelectric phase with peaks at $\SI{306}{\per\centi\meter}$ and $\SI{512}{\per\centi\meter}$. \textbf{c} Refractive index measurements with a prism coupler for a sol-gel derived barium titanate thin film on silicon before annealing only heating up to $\SI{400}{\celsius}$ (red) and after high-temperature annealing up to $\SI{700}{\celsius}$ on fused quartz (blue). In comparison, the extraordinary $(n_e)$ refractive index of bulk barium titanate is plotted (black).\cite{Wong1997}  \textbf{d} Spectroscopic second-harmonic generation measurement with the pump fs-laser wavelength at $\SI{880}{\nano\meter}$ and the generated second-harmonic at $\SI{440}{\nano\meter}$. The low intensity signal below $\SI{700}{\nano\meter}$ is from the setup (SI Figure S3). In the inset, the power-dependent SHG intensity (black dots) up to $\SI{80}{\milli\watt}$ follows a clear quadratic dependence on the fundamental power with the fitting function $8.47\cdot \mathrm{P}_{\mathrm{fund}}^2$ (red line).}
  \label{Fig:Characterization}
\end{figure}

Barium titanate is an established material in the electronics industry for applications requiring the piezoelectric effect.\cite{Acosta2017} In this context, different approaches for obtaining this material from sol-gel chemistry have been reported. Described methods comprise the formation of a mixed titanium-barium complex and heating aqueous suspensions of barium hydroxide and titanium dioxide in soft conditions or exploiting hydrothermal synthesis.\cite{Jolivet2000} 
Here, we use the sol-gel synthesis described in \cite{Edmondson2020}.\\
To study the crystalline and optical properties of the derived barium titanate, we prepared $\SI{680}{\nano\meter}$ thin films by ten iterations of sol-gel spin-coating and annealing (Methods). We confirmed by X-ray diffraction measurements (Methods) that the final material after annealing is indeed barium titanate (Figure \ref{Fig:Characterization}a) in agreement with previous studies.\cite{Tangwiwat2005} Raman spectroscopy measurements (Methods) are performed to demonstrate that barium titanate presents a distinct feature at $\SI{306}{\per\centi\meter}$ originating from splitting of molecular vibrational mode into longitudinal and transverse counterparts due to the tetragonal crystalline phase (Figure \ref{Fig:Characterization}b).\cite{Marssi03} The peak below $\SI{300}{\per\centi\meter}$ is due to the $\mathrm{A}_1(\mathrm{TO})$ phonon mode.\cite{Shiratori2007}
The sol-gel synthesis thus indeed results in a non-centrosymmetric tetragonal ferroelectric phase at room temperature and will therefore have a nonzero $\chi^{(2)}$-tensor.\cite{Edmondson2020} We measure the refractive index above the band gap at $\SI{387}{\nano\meter}$ of the deposited thin films with a prism coupling technique at at least three different wavelengths and extract a Cauchy fit from these (Figure \ref{Fig:Characterization}c).\cite{Karvounis2020, Ulrich1973} The material's refractive index depends on the underlying substrate and the annealing temperature. The substrate influences the crystallinity of the deposited material. We investigate the dependence of refractive index on annealing temperature and substrate, which show a refractive index increase post annealing indicating the densification of the resulting material. The highest refractive index values can be achieved by matching the lattice constant of BTO with the one of the underlying substrate, e.g., magnesium oxide.\cite{Tian2001,Cao2021} Finally, we perform spectroscopic and power-dependent second-harmonic generation measurements to confirm the second-harmonic nature of our material (Figure \ref{Fig:Characterization}d). The power-dependent SHG measurements follow a quadratic dependence on the fundamental power ($8.47\cdot \mathrm{P}_{\mathrm{fund}}^2$). The low-intensity signal below
$\SI{700}{\nano\meter}$ is from the setup (SI Figure S3). Therefore, the emission spectrum analysis confirms that our measured nonlinear signals are purely second-harmonic without fluorescence (Figure S3).
\subsection*{Nonlinear Periodicity in Three Dimensions}
The key feature of a three-dimensional nonlinear photonic crystal is a periodic nonlinear susceptibility in three dimensions.\cite{Slusher2003} Here, we investigate photonic crystals with a periodic variation of the second-harmonic generated signals. Generally, the second-harmonic periodicity may or may not coincide with the periodicity of the refractive index. In our structures, however, the spatial variation of the refractive index and the $\chi^{(2)}$ are expected to coincide. We investigate the nonlinear periodicity of our structure with a commercially available 2-photon-microscope in reflection (Methods). We excite with a fs-laser at $\SI{880}{\nano\meter}$ with a $25$x water immersion objective and detect the backscattered SHG signal passing through a $\SI{435}{\nano\meter}-\SI{455}{\nano\meter}$ filter onto an external non-descanned-detector (NDD) detector.
\begin{figure}
\centering
  \includegraphics[scale=0.75]{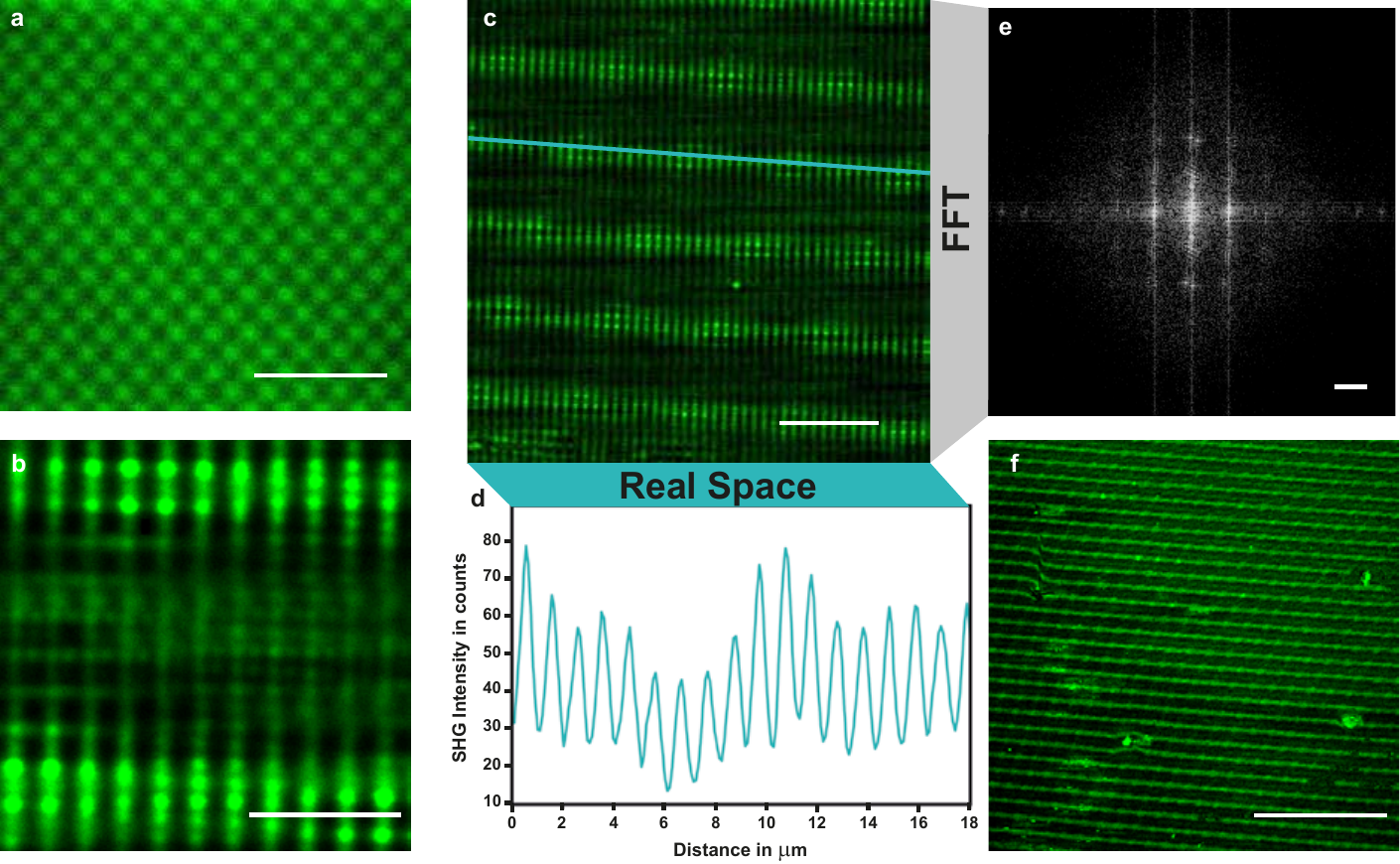}
  \caption{\textbf{Nonlinear Photonic Crystals} \textbf{a} 2-photon microscopy image depicting the periodic variation of the second-harmonic signal at $\SI{440}{\nano\meter}$ of a two-layer sample. Due to more material at the crossing points of the two layers, the crosses are brighter. \textbf{b} 2-photon microscopy image of a 4-layer woodpile sample. \textbf{c} 2-photon microscopy image of a larger area of the same 4-layer sample. Due to small rotations between consecutive layers, a Moiré pattern is visible in the nonlinear response. The cyan-colored line indicates the evaluation of the nonlinear period in real space \textbf{d}, which is $\SI{1}{\micro\meter}$ in agreement with the structural periodicity from SEM imaging. \textbf{e} A fast Fourier transform is applied to the image in \textbf{c}, and it reveals the pattern expected from a woodpile photonic crystal and determines the period to be $\SI{1}{\micro\meter}$ such that the stacking is of ABAB type. \textbf{f} 2-photon microscopy image of a large surface area of $5.3\cdot10^{4}~\SI{}{\micro\meter\squared}$ in a 4-layer sample. Scale bars are \textbf{a} $\SI{5}{\micro\meter}$, \textbf{b} $\SI{3}{\micro\meter}$, \textbf{c} $\SI{10}{\micro\meter}$, \textbf{e} $\SI{1}{\per\micro\meter}$ and \textbf{f} $\SI{75}{\micro\meter}$.}
  \label{Fig.4}
\end{figure}
First, we investigate the second-harmonic signals of a 2-layer sample (Figure \ref{Fig.4}a). The nonlinear optical periodicity in the $xy$ plane is clearly visible. The crossing points in the structures exhibit stronger second-harmonic signals due to the increased propagation length through the material.\cite{Boydbook2003} Figure \ref{Fig.4}b is a close-up of Figure \ref{Fig.4}c and shows a 2-photon microscopy image of a 4-layer sample. The green lines are the emitted signals from the rods of the woodpile photonic crystal. Due to small fabrication misalignment between the layers, a Moiré pattern is also present in the nonlinear signal (Figure \ref{Fig.4}c). In the areas where the spatial distance is smaller between the rods due to the misalignment, the overall second-harmonic signal is stronger. Imaging the SHG-intensity in real space (Figure \ref{Fig.4}d) reveals a nonlinear optical periodicity of $\SI{1}{\micro\meter}$ in agreement with the SEM analysis in Figures \ref{Fig.1} and \ref{Fig.2}. Applying a fast Fourier transform (FFT) algorithm onto the image in Figure \ref{Fig.4}c further confirms the woodpile photonic crystal geometry with a nonlinear optical periodicity of $\SI{1}{\micro\meter}$. This indicates that the geometry within the photonic crystal follows rather an ABAB-geometry than ABCD as observed in the SEM images of the structures (Figures \ref{Fig.2}a and c): If the geometry was mainly ABCD, the distance between the intensity maxima in the FFT would be doubled, i.e. $\SI{2}{\per\micro\meter}$ due to the reduced period of $\SI{0.5}{\micro\meter}$ originating from the half period shift for the C and D layers with respect to A and B. (SI Figure S4). Finally, Figure \ref{Fig.4}f shows a large-scale image of the fabricated nonlinear photonic crystals containing $5.3\cdot10^{4}$ unit cells which are 106 times (i.e., two orders of magnitude) more than the state-of-the-art. Resolving the second-harmonic periodicity in the $z$-plane is below the resolution limit of the optical system. However, the combined information from Figure \ref{Fig.4} and \ref{Fig.2} provides clear evidence that we indeed obtained 3D NPhCs.\\
Finally, we discuss the original phase-matching properties of our structures, which are linked to the different length scales in the system. The single building blocks of the woodpile are the individual rods with a cross-section of $150\times\SI{185}{\nano\meter\squared}$. They are composed of randomly oriented polycrystalline barium titanate domains. We recently demonstrated that the SHG efficiency within a random nonlinear medium can be described by random quasi-phase-matching and scales linearly with the number of domains, even if they are on the scale of nm.\cite{Savo2020} Furthermore, the average coherence length of $\ce{BaTiO3}$ at $\SI{930}{\nano\meter}$ is $\SI{1.23}{\micro\meter}$.\cite{Savo2020} Therefore, the building block dimensions are smaller than the coherence length of the material and do not suffer from phase mismatch. Though, on the length scale of one individual layer of rods, there is a periodic variation of an effective $\chi^{(2)}$ (Figure \ref{Fig.4}a). It is known that the SHG efficiency scales quadratically with the interaction length within a periodic nonlinear medium in the case of phase-matching and linearly for the quasi-phase-matching case.\cite{Boydbook2003,Zhang2021} In this work, the structures are 3D nonlinear photonic crystals with a periodic effective $\chi^{(2)}$ in three dimensions with a thickness below the coherence length and therefore no observable influence of phase-matching, which demonstrates a new class of 3D nonlinear photonic crystals. Currently, there is no theoretical work on the combination of the two effects and what happens to the phase and amplitude of the second-harmonic signal if a randomly oriented nonlinear medium is periodically structured.

\section*{Discussion and Outlook}
In our work, we solve one of the current major technical challenges in the field of 3D nonlinear photonic crystals, namely, to achieve precise and large-scale photonic structures.\cite{Zhang2021} By use of soft-nanoimprint lithography together with barium titanate sol-gel chemistry, we are the first to report a 3D bottom-up fabricated nonlinear photonic crystals. Our NPhC is a woodpile structure with up to 8 layers and nonlinear periodicity of $\SI{1}{\micro\meter}$ ($xy$-plane) and $\SI{300}{\nano\meter}$ ($z$-plane). The total surface area is more than $5.3\cdot10^{4}~\SI{}{\micro\meter\squared}$ corresponding to $5.3 \cdot 10^{4}$ unit cells in the $xy$-plane. This is two orders of magnitude larger than the state-of-the-art.\\
To further optimize the optical performance of the structures, research should continue towards exploring sol-gel derived materials with higher $\chi^{(2)}$-tensors. One short-term improvement would be to decrease the evaporation time of the sol-gel to leave more time for the alignment between layers. To achieve a large number of layers in the $z$-direction, the alignment procedure between consecutive layers has to be increased, for example, by using a commercial nanoimprint setup. Other improvements to increase the second-harmonic efficiency could comprise the electric poling of the material, the optimization of the duty cycle to $\SI{50}{\percent}$, and the addition of a dispersive material in the current air phase of the structure.\\
Investigating the nonlinear response of these structures at the band edges is a perfect test bed for fundamental physics, especially slow light enhancement of second-harmonic generated signals.\cite{Khurgin2023,Joannopoulos2007} Remarkably, in 3D NPhCs, depending on the collection angle and the incident wavelengths, either second-harmonic generation, sum-frequency or difference-frequency generation, as well as spontaneous parametric down-conversion (SPDC) can be observed.\cite{Zhang2021} The latter has, however, not been demonstrated experimentally to date. Theoretically, SPDC has only been studied in one-dimensional NPhCs.\cite{Vamivakas2004} The theory should therefore be extended to two and three dimensions. It is expected that several entangled photon pairs can be generated simultaneously due to the different phase-matching conditions in the different spatial directions of the photonic crystal.\cite{Keren-Zur2018} Another potential application could be cascaded frequency generation.\cite{Li2022} Further work in theoretical physics should address the influence of a polycrystalline periodic structure with respect to SHG amplitude, phase, and efficiency. Our novel method of fabrication provides an experimental platform to study this question. \\ 
We anticipate our approach to establish a new research area for bottom-up fabrication of 3D NPhCs. As a result, theoretically predicted effects can be studied, and ultranarrow multi-wavelength and coherent (quantum) light sources are envisioned. 

\section*{Methods}\label{Methods}
\subsection*{Preparation of Barium Titanate Sol-Gel}
The diol-based sol-gel synthesis approach that we use is based on the publication by Edmondson et al.\cite{Edmondson2020} We use barium acetate ($\ce{CH3COO)2Ba}$, ACS reagent $\SI{99}{\percent}$, Sigma Aldrich), acetic acid ($\ce{CH3CO2H}$, $\SI{99.7}{\percent}$, Alfa Aesar), acetylacetone ($\ce{CH3COCH2COCH3}$, Reagent Plus $\geq \SI{99}{\percent}$) and titanium(IV)isopropoxide ($\ce{Ti[OCH(CH3)2]4}$, $\SI{97}{\percent}$, Sigma Aldrich). The synthesis can be divided into three steps. First, a $\SI{0.2}{\mol\per\liter}$ solution of barium acetate in acetic acid is prepared. Therefore, $\SI{10}{\milli\liter}$ of acetic acid with $\SI{0.511}{\gram}$ of barium acetate are mixed, sealed with a parafilm, and stirred for one h. A beaker is placed inside the bath with a magnetic fish stirring slowly at $\SI{100}{\rpm}$. Then, first $\SI{426}{\micro\liter}$ of acetylacetate is added, followed by drop-by-drop titration of $\SI{592}{\micro\liter}$ of titanium(IV)isopropoxide. In the last step, the dissolved barium acetate is slowly added. The revolutions are increased to $\SI{300}{\rpm}$, and the Erlenmeyer beaker is closed with a parafilm. The reaction takes place for $\SI{24}{\hour}$. Afterward, the prepared sol-gel is diluted with $\SI{10}{\milli\liter}$ anhydrous methanol.

\subsection*{Preparation of PDMS Stamps}
The PDMS stamps are prepared from Sylgard 184 silicone elastomer kit from Dowsil. Base and curing agents are mixed in a $10:1$ ratio, and trapped air is removed by applying vacuum to the solution. The PDMS is poured on the silicon master mold, and bubbles are again removed by low pressure. The PDMS is hardened in the oven at $\SI{60}{\celsius}$ for at least three hours.

\subsection*{Solution for Planarization}
The planarization solution is a ($15~\mathrm{wt.}~\SI{}{\percent}$) solution of Norland Optical Adhesive (NOA60, Norland Products Inc.) with propylene glycol monomethylether acetate (PGMEA, Reagent Plus $ \SI{99.5}{\percent}$, Sigma-Aldrich). After spin coating, the films are exposed to UV light at an intensity of $\SI{150}{\milli\watt\per\square\centi\meter}$.

\subsection*{Thin Film Preparation}
 Barium titanate thin film for material characterization is prepared on plasma-treated fused quartz by spin-coating the sol-gel stock solution ($v=\SI{1000}{\rpm}$, $a=\SI{1000}{\rpm\per}$, $t=\SI{40}{\second}$) followed by annealing to $\SI{450}{\celsius}$ on a hot-plate. The spin-coating and annealing were iterated 10 times, and the final stack was annealed to $\SI{800}{\celsius}$ in a furnace with heating and cooling rate of $\SI{2}{\celsius\per\minute}$. The resulting film has a $\SI{680}{\nano\meter}$ thickness.

\subsection*{X-ray Diffraction Measurements and Raman Spectroscopy}
The crystallinity of the thin film annealed sol-gel BTO was characterized using standard Theta-2Theta XRD continuous rotating configuration  (Panalytical X'Pert PRO MRD with Ge monochromator and Cu $K_{\alpha}$ source at $\SI{40}{\kilo\volt}$, $\SI{45}{\milli\ampere}$). Raman spectroscopy measurements were acquired in an upright microscope setup with a 50x objective (Olympus LMPlanFLN), with a $\SI{532}{\nano\meter}$ laser source ($\SI{300}{\milli\watt}$, Nd: Yag laser Cobolt Samba) and a spectrometer ($\SI{5}{\percent}$ collection filter, grating $\SI{300}{\gram\per\milli\meter}$,  LabRAM HR Evolution UV-VIS-NIR, Horiba). 

\subsection*{Refractive index measurements}
The effective refractive index of the thin films was measured directly with commercial prism coupling setup (2010/M, Metricon) using wavelengths at $\SI{406}{\nano\meter}$, $\SI{526}{\nano\meter}$ and $\SI{638}{\nano\meter}$ to find at least two angular positions where guiding into the thin film occurred.\cite{Ulrich1973} This enabled us to directly infer the film's refractive index without the need to estimate the film thickness. Due to significant changes in the refractive index, low-temperature treated sol-gel films on Si substrate were measured for increased refractive index contrast. High-temperature treated films could be measured on fused quartz substrate due to sufficient contrast post-annealing. Tool software determined the Cauchy fit for the measured data points. 

\subsection*{Spectroscopic Second-Harmonic Measurements}
The thin film sol-gel BTO SHG was characterized using a Ti: Sapphire $\SI{120}{\femto\second}$ pulsed laser with a repetition rate at approximately $\SI{80}{\mega\hertz}$. Direct stray emission from the laser was minimized using a filter (FELH0650, Thorlabs) and is shown in SI Figure S3. to be the origin of the detected signal between 650-700 nm in Fig 3.d.  The incident average power on the sample was adjusted using a lambda-half plate (AHWP05M-980, ThorLabs),  Glen-Taylor polarizer  (GT10, ThorLabs), and a Si power detector (S132C, ThorLabs). The film was characterized in transmission configuration using a spherical lens (A240TM, $\mathrm{NA} = 0.5$, Thorlabs) with a $\SI{5}{\micro\meter}$ focused beam radius in the excitation through the substrate and a 100x objective (LMPlanFL N, $\mathrm{NA} = 0.8$, Olympus) for collection at the film side. The signal was focused on the spectrometer input port (Kymera 328i, grating groove density $\SI{150}{\liter\per\milli\meter}$, Andor) using two convex lenses (LA1461, ThorLabs) and a filter (FGB39M, ThorLabs) resulting in approximately $10^3$ times reduction of the fundamental laser power. The spectrum was acquired with a sCMOS camera (Newton 971, Andor) with 1 second exposure time accumulated over ten frames. 

\subsection*{2-Photon-Microscopy}
2-photon-microscopy imaging is performed on a commercial inverted Leica TCS SP8 setup equipped for multiphoton acquisition (Two-Photon Fluorescence excitation and Second-Harmonic Generation). We excite our structures at $\SI{880}{\nano\meter}$ with a Ti:Sapphire $\SI{100}{\femto\second}$ pulsed laser with a repetition rate of $\SI{80}{\mega\hertz}$ (Mai Tai XF, DeepSee (prechirp),  Newport Spectra-Physics) and regulated intensity via an electro-optic modulator. We use a Leica infrared apochromatic water immersion objective with $25x$ magnification and a numerical aperture of $0.95$ (HCX IRAPO). We collect backscattered SHG signal right behind the objective through an external NDD (Non-Descanned-Detector)  Leica Hybrid detector (HyD SP GaAsP) passing through a filter ($435$-$\SI{455}{\nano\meter}$, Semrock/AHF).  We confirm SHG signal at $\SI{440}{\nano\meter}$ spectrally beforehand using an internal descanned HyD detector.

\subsection*{Fast Fourier Transform}
FFT has been calculated with DigitalMicrograph\textregistered~Version 3.51.3720.0

\medskip
\textbf{Funding Sources}
This work was supported by the Swiss National Science Foundation Grant 179099 and  150609, the European Union’s Horizon 2020 research and innovation program from the European Research Council under the Grant Agreement No. 714837 (Chi2-nano-oxides) and No. 862346 (PolarNon). A.K. acknowledges financial support from the European Union's research and innovation Programme under the Marie Skłodowska-Curie Grant agreement No 801459, FP-RESOMUS and the Swiss National Science Foundation through the NCCR MUST. H.W. acknowledges financial support from the Physics Department at ETH Zurich.

\medskip
\textbf{Acknowledgements}
Viola V. Vogler-Neuling acknowledges ScopeM of ETH Zurich for providing instrumentation for SHG measurements. Furthermore, the authors acknowledge support from the operation team of the Binning and Rohrer Nanotechnology Center (BRNC) and the operation team of FIRST—Center for Micro and Nanoscience at ETHZ. The authors thank Dr. Matthias Saba for critically reading the manuscript.

\medskip
\textbf{Contributions}
V.V.-N. and R.G designed the experiment, R.F. fabricated and analyzed the samples with contributions from V.V.-N., Ü.-L.T. and H.W. V.V.-N., Ü.-L.T., J.W., P.B., V.W and J.L. developed the sol-gel material. Ü.-L.T. prepared the thin-films of sol-gel derived BTO for the materials characterization. Ü.-L.T. and V.V.-N. characterized the material. V.V.-N. and Ü.-L.T. performed the 2-photon microscopy analysis with contributions from J.K. R.F. performed the FDTD simulations with contributions from G.F. A.K. helped with come FIB cross-sections.
V.V.-N. and Ü.-L.T. wrote the manuscript with contributions from R.G and J.K.

\medskip
\textbf{Supporting Information} 
Supporting Information is available from the Wiley Online Library or from the author.

\bibliographystyle{plain}
\bibliography{main.bib}

\end{document}